\documentclass[transmag]{IEEEtran}
\usepackage{latexsym}
\usepackage{graphicx}
\usepackage{amsfonts,amssymb,amsmath}
\usepackage{xcolor}
\usepackage{comment}
\usepackage{float}
\usepackage{amsmath}
\usepackage{hhline}
\usepackage{booktabs}
\usepackage[normalem]{ulem}
\usepackage[tableposition=top]{caption} 

\usepackage{tikz}
\usepackage{pgfplots}
\usetikzlibrary{calc}
\usetikzlibrary{backgrounds}

\def\BibTeX{{\rm B\kern-.05em{\sc i\kern-.025em b}\kern-.08em \kern-.1667em\lower.7ex\hbox{E}\kern-.125emX}}
\pgfplotsset{compat=1.17}
\begin{document}

\title{Efficient Neuromorphic Signal Processing with Loihi 2}

\author{Garrick Orchard, E. Paxon Frady, Daniel Ben Dayan Rubin, Sophia Sanborn,\\Sumit Bam Shrestha, Friedrich T. Sommer, and Mike Davies

\thanks{Submitted 30 June 2021.}
\thanks{All authors are with Intel Labs, Intel Corporation, Santa Clara, CA, 95054 USA e-mail: mike.davies@intel.com.}%
}

\IEEEtitleabstractindextext{
\begin{abstract}
The biologically inspired spiking neurons used in neuromorphic computing are nonlinear filters with dynamic state variables\textemdash very different from the stateless neuron models used in deep learning. 
The next version of Intel's neuromorphic research processor, Loihi~2, supports a wide range of stateful spiking neuron models with fully programmable dynamics.
Here we showcase advanced spiking neuron models that can be used to efficiently process streaming data in simulation experiments on emulated Loihi 2 hardware. In one example, Resonate-and-Fire (RF) neurons are used to compute the Short Time Fourier Transform (STFT) with similar computational complexity but 47x less output bandwidth than the conventional STFT. In another example, we describe an algorithm for optical flow estimation using spatiotemporal RF neurons that requires over 90x fewer operations than a conventional DNN-based solution. 
We also demonstrate promising preliminary results using backpropagation to train RF neurons for audio classification tasks.
Finally, we show that a cascade of Hopf resonators\textemdash a variant of the RF neuron\textemdash 
replicates novel properties of the cochlea and motivates an efficient spike-based spectrogram encoder.
\end{abstract}

}

\maketitle

\section{INTRODUCTION}
\IEEEPARstart{I}{n} the language of signal processing, biological neurons are nonlinear time-varying filters that 
give rise to remarkable intelligence, energy efficiency, and computational speed when interconnected in great numbers.
While traditional signal processing has perfected the design and analysis of linear time-invariant systems, computing with nonlinear time-varying systems is far less mature.

The great successes of deep neural networks 
hint at the potential power of neural networks for nonlinear signal processing, but today's DNNs were not developed for that purpose and their characteristics have diverged far from biological principles.  Conventional artificial neuron models are vastly simplified compared to biology, replacing rich temporal dynamics with a point nonlinearity, such as ReLU. 
While time series processing can be supported by deep artificial neural networks, either with iterative matrix multiplication (RNNs and LSTMs) or by repeatedly processing sliding windows of vectorized data samples (CNNs and transformers), these approaches incur extra cost compared to assembling networks with units that each operate as  a causal, online filter.

Neuromorphic chips such as Intel’s Loihi~\cite{davies2018micro} implement temporal neuron models with dynamical behavior more similar to biological neurons, and their sparse communication and connectivity features support efficient scaling to high dimensional processing.
Loihi’s basic neuron model is a two-stage cascaded first order discrete time IIR filter producing output impulses (spikes) 
when sufficiently activated.  
This simple filter, when interconnected to thousands of other such filters, supports a rich space of nonlinear dynamics that can solve a wide range of problems, including graph search, similarity search, LASSO regression, and combinatorial optimization. 
In many cases, Loihi provides orders of magnitude gains in speed and energy compared to conventional solutions~\cite{davies2021ieee}.

Today the value of neuromorphic networks as signal processors remains underappreciated and underexplored. With Loihi 2, we have augmented Loihi with enhancements aimed at expanding the breadth of signal processing problems the architecture supports.
A full description of the architecture will follow in a future paper. 
This paper discusses some of the enhancements and shares early examples that showcase the 
value of Loihi 2's richer spiking neural network feature set for intelligent and efficient signal processing applications.

After introducing Loihi 2 in the next section, we demonstrate its richer feature set in simulation experiments following constraints of the hardware. By implementing Resonate and Fire (RF) neurons with oscillatory dynamics, we show how these neurons efficiently approximate the Short-Time Fourier Transform on audio signals.
We then show how RF neurons can be used in 
vision 
to estimate optical flow on event-data more accurately and efficiently than the standard EV-FlowNet model. 
We further demonstrate how networks of RF neurons can be trained with backpropagation and present early results on the NTIDIGITS and Google Speech Commands datasets.
Finally, we tweak the RF neuron model to create a Hopf Resonator and describe its use in cochlea modelling.

\begin{figure*}[h]
\centerline{\includegraphics[width=\textwidth]{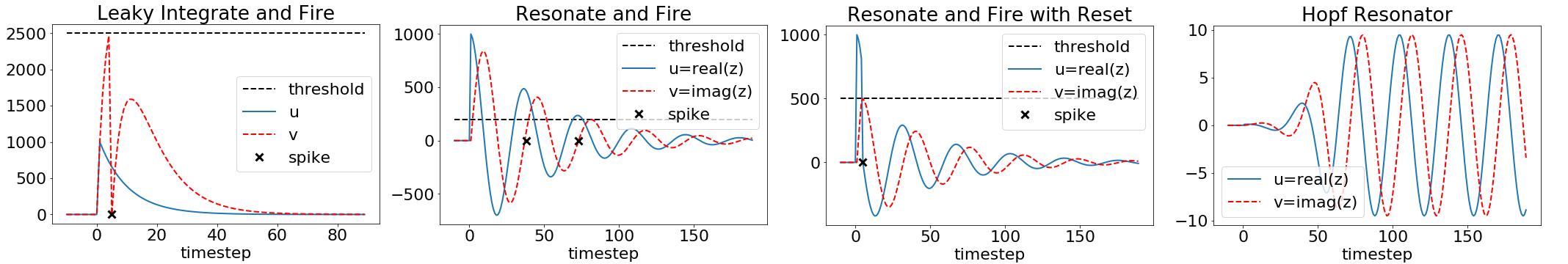}}
\caption{Response of neuron models used in this paper to an impulse (spike) at time 0. a) A Leaky Integrate and Fire model which spikes whenever voltage exceeds threshold. b) A complex valued Resonate and Fire model used for spatiotemporal filtering, which spikes whenever $z$ crosses the real axis and $real(z)$ is greater than threshold. c) A complex valued Resonate and Fire model with reset used in deep networks. d) A  complex valued Hopf Resonator with a stable limit cycle.
\label{fig:neuron_models}}
\end{figure*}
\begin{figure}[h]
\centerline{\includegraphics[width=0.47\textwidth]{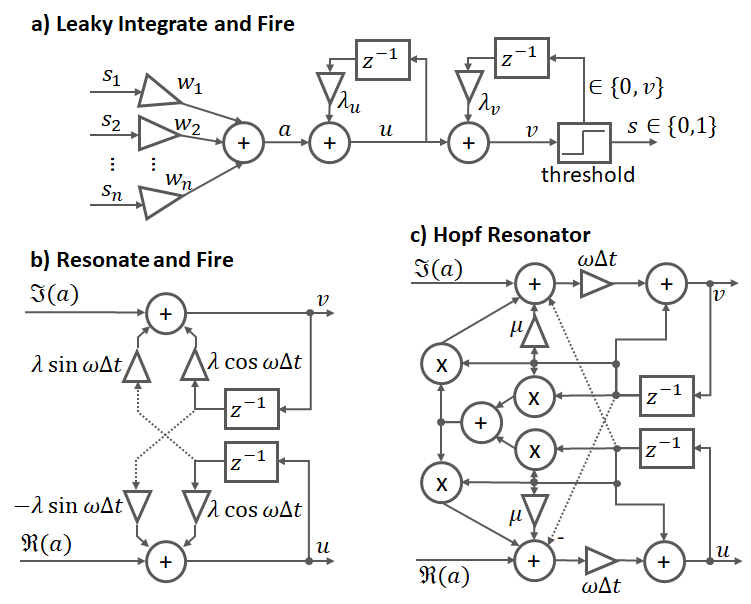}}
\caption{Block diagrams of the discrete computation for the LIF neuron, and RF and Hopf resonators.
The Hopf Resonator shows an implementation of the Euler method, but in practice we use a 4th order Runge-Kutta.
\label{fig:z_diagram}}
\end{figure}

\section{Loihi Architecture}
Spiking neurons have been modelled as first order differential equations since Hodgkin and Huxley. 
Early Neuromorphic Engineers mimicked the dynamics of biological spiking neurons in silicon using analog circuit dynamics  \cite{arthur2010tcsi}.
More recently, there has been a shift towards digital implementations, both in software and in silicon, which naturally gives rise to a discrete time formulation of the internal neuron dynamics
Both Loihi and its successor, Loihi~2, use this digital approach.

Loihi implements the discrete Leaky Integrate and Fire (LIF) neuron model
\begin{align} \label{eq:lif}
    a_i[t] = \sum_j w_{ij}s_j[t-1] \\
    u_i[t] = \lambda_uu_i[t-1] + a_i[t]\\ 
    v_i[t] = \lambda_vv_i[t-1] + u_i[t] 
\end{align} 
where $a_i[t]$ is the accumulated synaptic activation for timestep $t$, $u_i$ and $v_i$ represent the $i^\text{th}$ neuron's current and voltage respectively, and $\lambda_u$ and $\lambda_v$ are the current and voltage decay. Whenever $v_i[t]$ exceeds threshold, a spike is generated ($s_i[t]=1$) and the voltage variable is reset to zero ($v_i[t]=0$). All states and parameters use fixed precision.
Loihi 2 introduces a more flexible microcode programmable neural engine. 
Users can allocate variables and execute a wide range of instructions organized as short programs using an assembly language. 
These programs have access to neural state memory, the accumulated synaptic input $a_i$ for the present timestep, random bits for stochastic models, and a timestep counter for time-gated computation. 
The instruction set supports conditional branching, bitwise logic, and fixed-point arithmetic backed by hardware multipliers.

Within a core, memory limits the number of different neurons which can be implemented.
By using lower precision neuron models, more neurons can be implemented within the same memory footprint, up to a maximum of 8192 per core.
More complicated neurons can be implemented as longer programs which access multiple memory addresses for neural state and synaptic input, and pass information to each other through the persistent thread state.

In addition to allowing much richer internal neuron dynamics, the neural engine allows user-defined output nonlinearities and reset mechanisms, and can generate 32-bit graded spikes. 
These are all new features in Loihi 2 which we make use of in this paper. 
Fig.~\ref{fig:neuron_models} shows the impulse (spike) response of the different Loihi~2 neuron models described in this paper, including their spiking output and reset behavior.
Fig.~\ref{fig:z_diagram} shows the computation associated with these neuron models. While the LIF model uses a two stage cascade of filters, the RF and Hopf Resonator models use two cross-coupled filters to generate the real and imaginary components.

Loihi 2 further provides richer connectivity features than its predecessor. Synaptic activations can be computed from graded spikes, support for convolutional connections has been optimized, and new features allow procedural generation of stochastic synapses and separable synaptic matrices.

\begin{figure}[t]
\centerline{\includegraphics[width=0.45\textwidth]{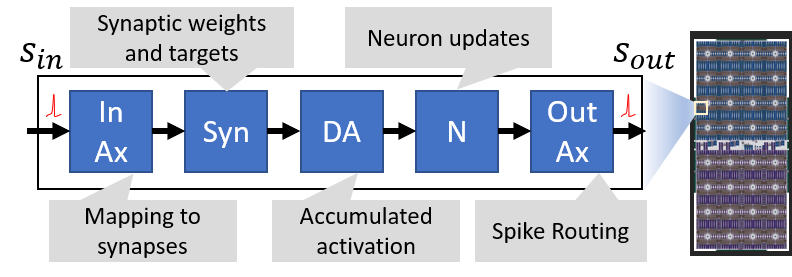}}
\caption{The Loihi 2 chip plot (right) and processing flow for a single core (left). Incoming spikes are mapped to lists of synapse weights which are accumulated for consumption in the next timestep. Meanwhile, neurons update using the previous timestep's accumulated activation and generate spikes which are routed to other cores by the Output Axon stage.
\label{fig:coreflow}}
\end{figure}
\begin{figure*}[ht!]
\centerline{\includegraphics[width=\textwidth]{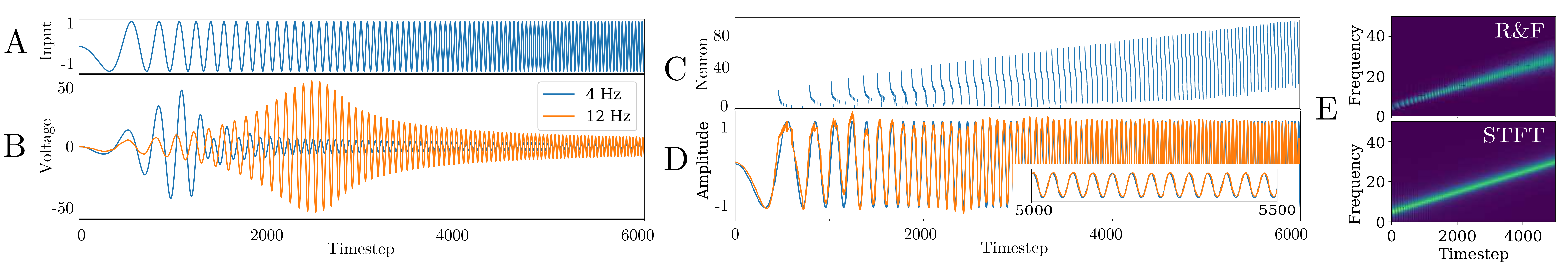}}
\caption{Using simulated RF neurons compatible with Loihi 2 to approximate the STFT of an audio chirp and reconstructing the input signal from the spike responses. A: The input chirp. B: Real component of the internal dynamics of two RF neurons with different resonant frequencies. C: Spike raster showing the output timing for all 100 neurons in response to the stimulus. D:  Reconstruction (orange) of the original signal (blue) E: Comparison to the STFT.
\label{fig:RF_fourier}}
\end{figure*}

\section{Applications}

\subsection{Resonate-and-Fire Neurons for Spectral Analysis} \label{sec:app:rf_stft}

The Resonate-and-Fire (RF) neuron is an extension of the standard LIF model, newly enabled in Loihi 2.  The RF neuron is a damped harmonic oscillator with a spiking mechanism. Each neuron has a resonant frequency $\omega$, a complex-valued state $z = u + \mathrm{i}v$, and a decay factor $\lambda\in(0,1)$, with dynamics defined by
\begin{equation}\label{eqn:rf}
z_k[t] = \lambda e^{\mathrm{i}\omega\Delta t}z_k[t-1] + a_k[t]
\end{equation}
where the last term is the synaptic input and $\lambda e^{\mathrm{i}\omega\Delta t}$ defines the oscillation kernel. When starting from an initial condition of $z_k[t]=0$ and assuming no reset mechanism, the equation can be rewritten in the form
\begin{equation}
z_k[t] = \sum_n{e^{\mathrm{i}n\omega\Delta t}\lambda^na_k[t-n]}
\end{equation}
which is recognizable as the term for frequency $\omega$ of the discrete Short-Time Fourier Transform (STFT) of $a_k[t]$ with an exponential window. By integrating input on each RF neuron, the exponentially decaying sliding window naturally arises without needing to store a history of input samples. A bank of RF neurons at different frequencies can then be used to estimate the STFT.

A key distinguishing feature between a spiking neuron and a digital IIR filter is the neuron's temporally sparse pulsed output. For RF neurons, a typical spiking mechanism generates spikes at zero crossings of the imaginary part and if the real part surpasses a given threshold (Fig.~\ref{fig:neuron_models}b). With binary spikes, this neuron model can be used in conjunction with spike-timing codes to perform complex matrix algebra as shown in \cite{frady2019robust}. More generally, we may use graded spikes in Loihi 2 to transmit the magnitude $|z|$, which conveniently equals the real component $\mathfrak{Re}(z)$ when $z$ crosses the real axis.

By encoding a signal's spectrum in a sparse, event-driven manner with spikes, the communication bandwidth is automatically compressed without increasing latency. 
In the example shown in Fig.~\ref{fig:RF_fourier}, the RF implementation reduces output bandwidth by 47x compared to a conventional STFT producing a spectrogram vector on each time step.

We used RF neurons to produce STFTs for several examples in the Google Speech Commands dataset. 
The Spiking STFT produced by the RF neurons can be inverted by convolving the spikes with the neuron's oscillation kernel and integrating across the population. 
We varied the spiking threshold of the RF neurons and measured the reconstruction correlation as a function of spikes produced. We compared this to the conventional STFT by excluding the smallest coefficients and performed inverse STFT to measure the reconstruction correlation. 
Reconstructions from the RF-generated spikes saturate to 94\% correlation with only five thousand spikes. A conventional STFT computation generates over 3 million complex values over the same period. Computing a reconstruction from only the largest 500K nonzero coefficients maintains a high reconstruction correlation of 98\%, but the reconstruction correlation drops to $\sim$63\% 
if only the largest 5,000 values are preserved. 
This shows that the RF implementation uses spike timing to encode the STFT information more efficiently than the conventional STFT.

\begin{figure}
    \centering
    \includegraphics[width=0.47\textwidth]{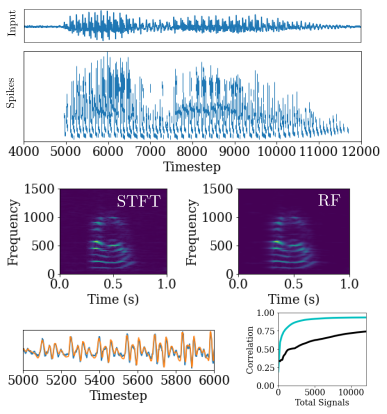}
    \caption{Spiking STFT for google speech command (top). Conventional STFT is shown compared to RF based STFT (middle). Reconstruction algorithm reproduces original signal (bottom). Reconstruction efficiency of RF model (cyan) is compared to thresholded conventional STFT (black). }
    \label{fig:speech_stft}
\end{figure}

\subsection{Resonate-and-Fire Neurons for Estimating Optical Flow}
\begin{figure}
\centerline{\includegraphics[width=0.45\textwidth]{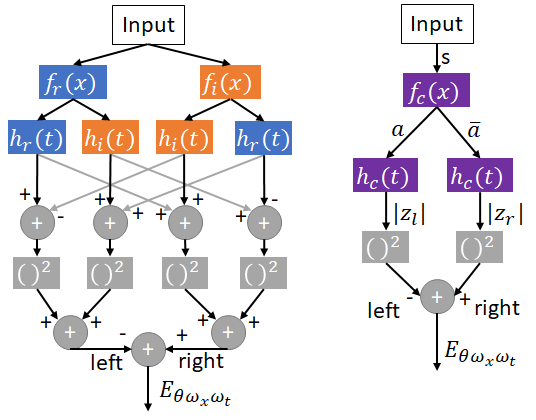}}
\caption{The original Adelson and Bergen Opponent Energy model (left) consisting of separable spatial $f()$ and temporal $h()$ components with orange filters out of phase with the blue filters. The equivalent model implemented using complex (purple) spatial and temporal filters.
\label{fig:energy_model}}
\end{figure}

While the membrane dynamics of a spiking neuron can implement a temporal filter, the synapses to a neuron can implement a spatial filter. 
By combining the two we can implement separable spatiotemporal filters.
The accumulated synaptic activation $a_i$ in \eqref{eq:lif} holds the result of the spatial filtering and can be shared in Loihi 2 between multiple neurons with different resonant temporal frequencies, allowing a single spatial filtering result to be re-used in multiple 
filters. 

Spatiotemporal filtering of this type is especially efficient when processing sparse spiking data 
produced by other spiking neurons, or by Dynamic Vision Sensors \cite{gallego2020pami}. Here we demonstrate how RF filters can be used to estimate optical flow using motion energy models \cite{adelson1985osa}. Recent work \cite{peveri2021cvprw} proposes a similar energy based approached based on LIF neurons. 

Fig.~\ref{fig:energy_model} left shows the 
opponent energy model from \cite{adelson1985osa}, consisting of separate spatial $f()$ and temporal $h()$ filters, each with either even (blue) or odd (orange) symmetry, which can also be represented using real and imaginary filter components. Fig.~\ref{fig:energy_model} right shows an equivalent complex model using RF neurons which can operate on either conventional image frames or sparse event-based DVS data. In the case of frames, a separate graded spike is used to represent the intensity of each pixel. For DVS data the spikes are used directly.

Spikes, $s$, feed into a layer of complex synapses which result in a complex accumulated synaptic activation $a$ for every timestep.
The activation $a$ and its conjugate $\bar{a}$ feed into two identical RF neurons which output the magnitude of their activation once per oscillation using the graded spiking mechanism described in Section~\ref{sec:app:rf_stft}. The spike mechanism guarantees that the imaginary component is at or near 0 and can be ignored, so only the real components need to be squared and summed to compute the opponent energy. A single Loihi~2 neuron program can implement both neurons as well as the square and summation to directly output the opponent energy as a graded spike.

The model in Fig.~\ref{fig:energy_model} only computes the opponent energy $E$ for one combination of orientation $\theta$, spatial frequency $\omega_x$, and temporal frequency $\omega_t$. The opponent energies from multiple filters of different frequencies and orientations must be combined in post-processing off-chip to arrive at an optical flow estimate. Each neuron with spatial frequency $\omega_x$, temporal frequency $\omega_t$ and orientation $\theta$ has a preferred input velocity $v \in \mathbb{R}^2$, which is orthogonal to the orientation of the neuron's spatial receptive field.
\begin{align} 
    v_{\omega_x,\omega_t, \theta} &= \begin{bmatrix}
           \frac{\omega_t}{\omega_x}\cos{\theta}, & 
           \frac{\omega_t}{\omega_x}\sin{\theta} \\
         \end{bmatrix} \label{eq:velocitycomponents}
\end{align}

Stimuli moving at the neuron's preferred velocity yield the highest magnitude inner product with its spatiotemporal receptive field. 
To estimate the optical flow $f \in \mathbb{R}^2$ at a given pixel location, the neurons' preferred velocities are weighted by their normalized opponent energy at that pixel location.

\begin{align}  \label{eq:flownormalization}
    f &= \frac{\sum\limits_{\omega_x,\omega_t, \theta}v_{\omega_x,\omega_t, \theta}E_{\omega_x,\omega_t, \theta}}{\sum\limits_{\omega_x,\omega_t, \theta}{E_{\omega_x,\omega_t, \theta}}}
\end{align} 

\begin{table}[t!]
    \centering
    \caption{Optical flow model parameters}
    \begin{tabular}{l|c|c|c|c}
    \textbf{Parameter} & \textbf{Units} & \textbf{Symbol} & \textbf{Count} & \textbf{Values} \\ \hline 
    Receptive Field Size & pix & - & - & $(64, 64)$ \\ 
    Timestep Duration          & sec        & $\Delta t$      & - & 0.032 \\
    Spatial Frequency  & rad/pix        & $\omega_x$      & $n_x=1$ & $\omega_{x}=\frac{6\pi}{256}$\\
    Temporal Frequency  & rad/sec   & $\omega_t$      & $n_t=5$ & $\omega_{t_k}=4\pi k$\\
    Orientations        & rad       & $\theta$        & $n_{\theta}=4$   & $\theta_{k} = \frac{k\pi}{n_{\theta}}$\\
    \end{tabular}
    \label{tab:OFparams}
\end{table}

We implement spatiotemporal filters with Gabor shaped receptive fields of different orientations using the parameters in Table~\ref{tab:OFparams} to estimate optical flow from event data.
We evaluate our model on the Multi View Stereo Event Camera (MVSEC) dataset \cite{zhu2018ral} and compare our results to those obtained by EV-FlowNet, a state-of-the-art model for estimating optical flow from event data. 
EV-FlowNet is a deep stateless neural network trained under self-supervision. 
Our model, by contrast, requires no training data, and processes event data timestep by timestep instead of buffering input data and presenting it statically as voxels to a deep network.

We compare to the 2R variant of EV-FlowNet, as it achieves the best performance across the test sequences in \cite{zhu2018ral}. 
We evaluate the models using the methods described in \cite{zhu2018ral}, calculating the Average Endpoint Error (AEE) and outlier percentage (percentage of flow vectors with AEE $> 3$ pixels).

\begin{table}[t]
\centering
\caption{Average Endpoint Error on MVSEC}
\resizebox{\linewidth}{!}{%
\begin{tabular}{l|ll|ll|ll}
   & \multicolumn{2}{l|}{\textbf{\vtop{\hbox{\strut Indoor}\hbox{\strut Flying 1}}}} & \multicolumn{2}{l|}{\textbf{\vtop{\hbox{\strut Indoor}\hbox{\strut Flying 2}}}} & \multicolumn{2}{l}{\textbf{\vtop{\hbox{\strut Indoor}\hbox{\strut Flying 3}}}} \\
 &  AEE  &   \% outlier &   AEE & \% outlier & AEE   & \% outlier \\ \hline
$\mbox{EV-FlowNet}_{2R}$ & 1.03 & 2.2 & 1.72 & 15.1 & 1.53 & 11.9 \\
Ours$_\text{DENSE}$ & 0.91 & \textbf{0.35} & 1.28 & 5.83 & 1.04 & 2.88 \\
Ours$_\text{SPIKES}$ & \textbf{0.83} & 0.68 & \textbf{1.22} & \textbf{5.42} & \textbf{0.97} & \textbf{2.65}
\end{tabular}}
\label{tab:OFresults}
\end{table}

\begin{figure}[t!]
\centerline{\includegraphics[width=0.5\textwidth]{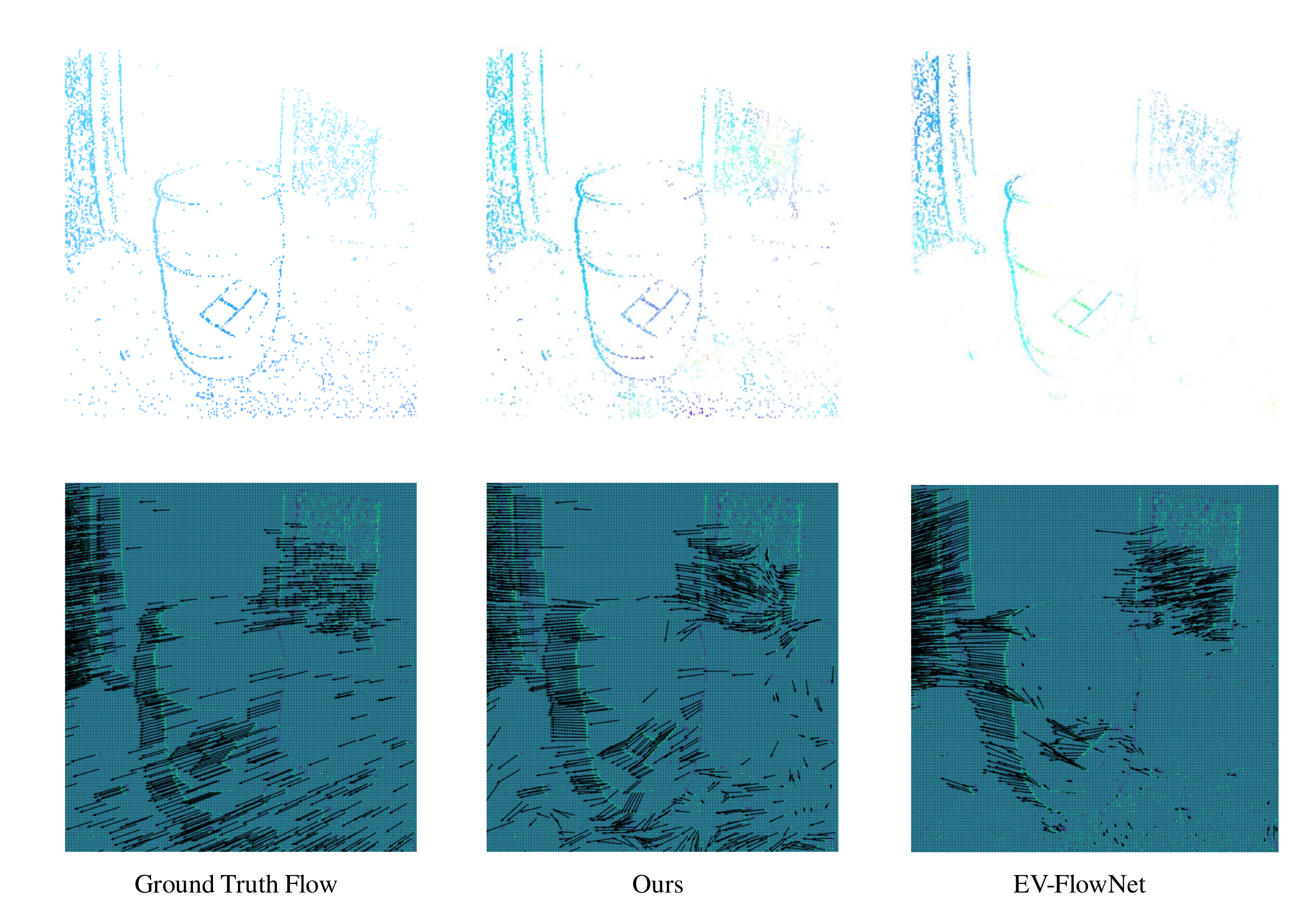}}
\caption{Ground truth optical flow from the MVSEC dataset (left) compared to our method (middle) and EV-FlowNet (right). 
\label{fig:optical_flow}}
\end{figure}

Table~\ref{tab:OFresults} compares results for two versions of our model.
The \emph{dense} version estimates flow directly from the neuron's internal state. 
The \emph{spikes} version uses the most recently received graded spike value from the neuron as its activation when estimating optical flow.
Across the three indoor flying sequences from \cite{zhu2018ral}, our models achieves better performance than EV-FlowNet on these two metrics.
Fig.~\ref{fig:optical_flow} illustrates a single representative frame.

Compared to EV-FlowNet, our model has less than half the neurons, although each RF neuron update is more expensive (4 MACs versus 1 ReLU). 
However, computation in EV-FlowNet is dominated by synops, which outnumber the neuron updates by 2000x.
Our model has over 10x fewer synapses, our synops are cheaper (a complex AC versus a MAC), and sparse synapse activation results in 93x fewer synops on the MVSEC sequences tested.

To compute optical flow from the neuron state, the neuron must square its magnitude to compute the opponent energy (Fig.~\ref{fig:energy_model}), and the energies must be combined in post-processing off-chip. 
Off-chip post-processing \eqref{eq:velocitycomponents} introduces another 2 MACs per neuron and one inversion per pixel, which is still dwarfed by the synaptic ops. 
For comparison, readout neurons in EV-FlowNet require computing $\tanh$ twice per pixel.

The RF optical flow model benefits from two key properties. 
The event-based operation exploits sparsity of the input spike data to reduce synaptic ops, and an overlapping region between two subsequent temporal windows only needs to be processed once, saving further ops. 

\subsection{Using Backpropagation to Train RF Neurons}
The success of deep learning comes from its ability to train large networks end-to-end with data, thereby avoiding laborious hand engineering.
To that end, we have extended the Spike Layer Error Reassignment (SLAYER) tool \cite{shrestha2018nips} that was used to train LIF neural networks for Loihi to handle complex and oscillatory models with graded spikes, including the RF neuron, for Loihi 2.

The extension to SLAYER tackles the temporal error credit assignment problem by redistributing the error by applying the decaying rotation operator $\lambda e^{-\mathrm{i}\omega\Delta t}$ back in time. 
For deep networks, we introduce a different output linearity for the RF neuron. Following the model proposed in \cite{izhikevich2001nn}, the RF neuron generates a unary spike whenever its imaginary component exceeds threshold, following which the real component is reset to 0 (Fig.~\ref{fig:neuron_models}c).
Approximation of the derivative of the spike function follows the relaxation of the derivative of spike threshold mechanism using nascent delta approximation.

Using SLAYER we trained a hybrid MLP of RF and LIF neurons (64-256RF-256RF-242LIF) on the spiking NTIDIGITS \cite{anumula18fns} audio dataset. 
The model, with 226K parameters, predicts digit utterances with an accuracy of $92.14\pm0.24\%$. 
In contrast, the best-known conventional solution using LSTM units (643K parameters) achieves an accuracy of $91.25\%$ \cite{anumula18fns}.

We also tackled the more challenging Google Speech Commands dataset \cite{warden2018arxiv} (10+2 subset) with both MLP and convolutional architectures by first converting the dataset to spikes using a publically available cochlea model~\cite{zilany2014}. 
The MLP architecture (64-256RF-256RF-288LIF) has 238K parameters and achieves an accuracy of $88.97\%$. 
An equivalent LIF only architecture with 156K parameters achieved an accuracy of $88.03\%$. 
A hybrid CNN architecture with five RF convolution layers followed by two LIF dense layers, achieves $91.74\%$.

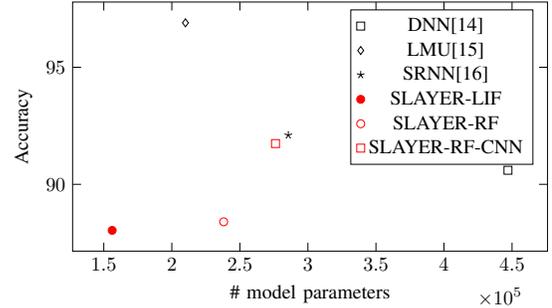
\begin{figure}[t]
\centering
\begin{tikzpicture}[scale=0.75, every node/.style={transform shape}]
\begin{axis}[
    height = 6cm,
    width = 10cm,
    tick pos=both,
    xtick style={color=black},
    ytick style={color=black},
    xlabel={\# model parameters},
    ylabel={Accuracy},
    legend pos=north east,
    only marks,
    xtick scale label code/.code={$\times 10^5$},
]
    \addplot[mark=square] coordinates {(447000, 90.6)}; 
    \addplot[mark=diamond] coordinates {(210000, 96.9)}; 
    \addplot[mark=star] coordinates {(285600, 92.1)}; 
    \addplot[mark=*, red] coordinates {(156160, 88.032)}; 
    \addplot[mark=o, red] coordinates {(238080, 88.398)}; 
    \addplot[mark=square, red] coordinates {(276176, 91.74)}; 
    \legend{
        DNN\cite{rybakov2020}, 
        LMU\cite{blouw2020arxiv},
        SRNN\cite{yin2021arxiv},
        SLAYER-LIF, 
        SLAYER-RF,
        SLAYER-RF-CNN,
    }
\end{axis}
\end{tikzpicture}
\caption{Performance comparison on Google Speech Commands 10+2 subset task.}
\label{fig:SubsetSCResults}
\end{figure}

The performance of SLAYER trained networks with other existing streaming models is shown in Fig.~\ref{fig:SubsetSCResults}. 
All of our SLAYER models are trained taking the fixed precision of Loihi 2 into account. 
The LMU \cite{blouw2020arxiv} and DNN datapoints use more conventional dense matrix vector arithmetic than the SLAYER model which naturally exploits and generates sparse data.
The LMU and SRNN \cite{yin2021arxiv} datapoints are also recurrent architectures whereas the SLAYER trained models presented here are feed-forward networks. 
To our knowledge, this is the first time backpropagation training has been demonstrated on a complex network of RF neurons to solve standard benchmarked problems, and we expect results to improve as we continue to explore novel and recurrent architectures.

\subsection{Extension to Cascaded Hopf Resonators}

A second-order nonlinearity added to the RF membrane dynamics  
results in a Hopf resonator, which is also supported by Loihi 2's programmable architecture. 
Its name comes from the Hopf bifurcation, which is a critical point at which a periodic solution to the differential equation arises because the system stability switches (Fig.~\ref{fig:neuron_models}d).
The distance to the critical point is adjusted by input strength $a$, which offers self-adjusting gain and bandwidth control. 

The Hopf dynamics have
wide-ranging signal processing applications, a classical one being regenerative receivers, which have recently seen a revival in low-power wireless applications \cite{tapson2008iscas}. 
Another prominent role of Hopf resonators is as active elements in models of the auditory pathway up to the auditory nerve \cite{eguiluz2000phyr,kern2003phyr}, which provides a class of biologically inspired methods of audio pre-processing. 
We find the approach based on Hopf resonators particularly convenient because of its simplicity, compactness and the many emerging properties it highlights.

The basal membrane in the cochlea oscillates, selectively amplifying audio frequency (ordered from base to apex - HF to LF) components.
In cochlea models \cite{eguiluz2000phyr}, membrane sections for different frequencies are modeled by a cascade of Hopf resonators.
Specifically, each membrane section provides a band-passed filtered version of its input to the following lower frequency section, and so on.

We model a cochlea section using the continuous formulation
\begin{equation}\label{eqn:hopf}
\dot z = \omega_0((\lambda-|z|^2 + i)z + a),
\end{equation}
where $z \in \mathbb{C}$ is the resonator response, and $\omega_{0},\;\lambda \in \mathbb{R}$ are, respectively, the characteristic frequency of the cochlear section and the distance from the resonating actual frequency at which the system diverges, while $a \in \mathbb{C}$ is the external input to the resonator. In this model we do not use a spiking output.
The choice of $\omega_0$ to scale the differential equation is more stable over the human tuning response characteristics and more conveniently requires normalized units of the forcing input and the $\lambda$ parameter \cite{kern2003phyr}.
On Loihi \eqref{eqn:hopf} is discretized using a 4th order Runge-Kutta method (Fig.~\ref{fig:neuron_models}d). 
\begin{figure}
\centerline{\includegraphics[width=0.5\textwidth]{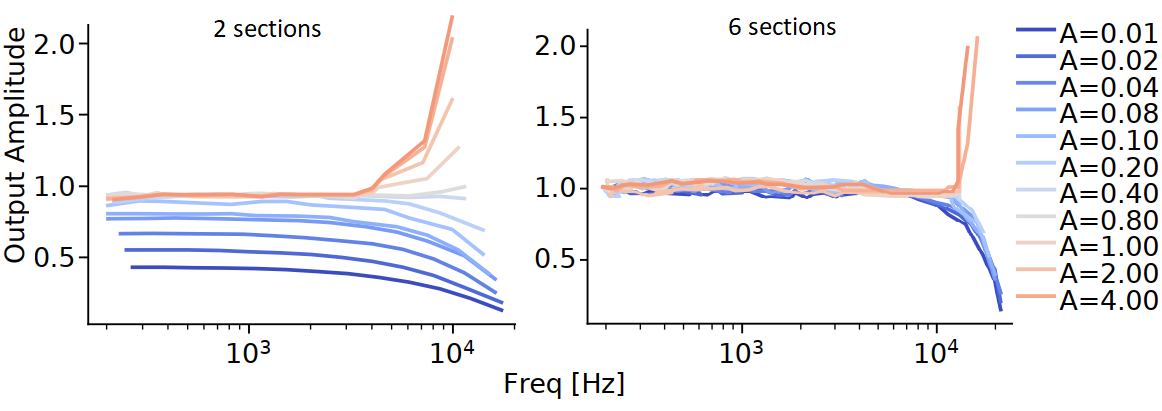}}
\caption{Amplitude normalization using a cascade of Hopf resonators. Peak output amplitude (y-axis) for a cascade of Hopf resonators plotted against input stimulus frequency (x-axis)  for different input amplitudes (A).  Increasing the number of sections per octave from 2 (left) to 6 (right) increases the sharpness of the normalization tuning. 
\label{fig:RF_hopf}}
\end{figure}

Following \cite{kern2003phyr}, each cochlear section is modeled by combining a Hopf resonator \eqref{eqn:hopf} with a 6-th order butterworth low pass filter with cutoff at $1.05\omega_0$. We have explored 
different densities of sections per octave (Fig.~\ref{fig:RF_hopf}) and demonstrate that, at higher densities, the cascading provides a \textit{self-normalizing gain} control. Any signal amplitude across several orders of magnitude of frequency range is normalized to a narrow dynamic range (-3dB, 0dB).
This self-normalizing gain control is a novel observation, an emergent property of cascading Hopf nonlinearities at certain densities. 
On Loihi 2, this cascade of resonators, projected onto a LIF downstream neuron, can provide a highly efficient spike encoder invariant to input peak amplitudes.

\section{Conclusion}
Loihi 2's generalized feature set, including a programmable neuron engine, provides far greater flexibility than Loihi for exploring novel spiking neural network models. 
In this paper we have shown how Loihi 2 can implement complex-valued oscillatory neurons, a natural first step beyond the much-studied leaky-integrate-and-fire model. We showed that these neurons can be used to approximate the STFT of a signal and compute the optical flow of visual data with significant savings in computational cost compared to conventional approaches. We have shown that these networks can be trained to recognize speech commands and can replicate the emergent signal processing features of the cochlea, highlighting promising directions for future research.
These represent a new class of computational tools for optimizing the energy, latency, and model sizes of intelligent signal processing applications when mapped to a neuromorphic architecture such as Loihi 2.

\bibliographystyle{IEEEtran}
\bibliography{bib/abbrv,bib/all,bib/rf_stft}
\end{document}